\documentclass[11pt]{article}
\usepackage[latin1]{inputenc}
\usepackage{latexsym}
\usepackage{amstext}
\usepackage{amsxtra}
\usepackage{epsf}

\newcommand{\valass}[1]{\left|#1\right|}
\newcommand{\norme}[1]{\left\|#1\right\|}
\newcommand{\ket}[1]{ \left. | #1 \right\rangle } 
 
\newcommand{\braket}[2]{ \left\langle #1 | #2 \right\rangle } 
\newcommand{\braopket}[3]{ \left\langle #1 | #2 | #3 \right\rangle } 
\newcommand{\ketbra}[2]{ | #1 \left\rangle \right\langle #2 |}
\newcommand{\mean}[1]{\overline{#1}}
\newcommand{\expect}[1]{\left\langle #1 \right\rangle}
\newcommand{\eq}[1]{(\ref{eq:#1})}
\newcommand{\al}[1]{^{(#1)}}
\newcommand{\Psihd}{\hat\Psi^\dagger}

\newcommand{\Psih}{\hat\Psi}

\begin{document}

%\draft
\title{The $N$ boson time dependent problem: \\
an exact approach with stochastic wave functions}

\author{I. Carusotto$^{(a,b)}$, Y.Castin$^{(a)}$ and J.Dalibard$^{(a)}$ \\
(a) Laboratoire Kastler-Brossel, Ecole Normale Sup\'erieure, 24
rue Lhomond, \\ 75231 Paris Cedex 05, France \\
(b) Scuola Normale Superiore and INFM, Piazza dei Cavalieri 7, 56126 Pisa, Italy}
\date{}
\maketitle

\begin{abstract}
We present a numerically tractable method to solve exactly the evolution
of a $N$ boson system with binary interactions. The density operator of the system
$\rho$ is obtained as the stochastic average of particular operators $|\Psi_1\rangle
\langle \Psi_2|$
of the system. The states $|\Psi_{1,2}\rangle$
are either Fock states $|N:\phi_{1,2}\rangle$ or 
coherent states $|\mbox{coh}:\phi_{1,2}\rangle$ with each particle in the state $\phi_{1,2}(x)$.
We determine the conditions on the evolution of $\phi_{1,2}$
--which involves a stochastic element-- under which we recover the exact evolution 
of $\rho$. 
We discuss various possible implementations of these conditions. The well known 
positive $P$-representation arises as a particular case of the coherent state 
ansatz. We treat numerically two examples: a two-mode system and a one-dimensional
harmonically confined gas. These examples, together with an analytical estimate of the
noise, show that the Fock state ansatz is the most promising one in terms of 
precision and stability  of the numerical solution.
\end{abstract}

\noindent {\bf Pacs:} 03.75.Fi, 05.10.Gg, 42.50.-p 

\parskip 2mm

%\narrowtext
\section{Introduction}

Since the experimental realization of the first atomic gaseous Bose-Einstein
condensates a few years ago \cite{Anderson95,Bradley957,Davis95,Fried98}, 
the physics of dilute Bose gases has
been considered with a renewed interest. One fascinating aspect of these
new systems is the possibility to accumulate in a single quantum state a large
fraction of the atoms confined in a trap. 

At very low temperature, a
simple theoretical description of the dynamics 
of these systems is obtained by neglecting
the uncondensed atoms, and by considering the 
wave function of the condensate,
which obeys a Schr\"odinger equation with a non 
linear term originating from
the mean-field interactions between the atoms.
Such an approach neglects two- and more-particle correlations and 
is valid under a weak-interaction condition which is usually stated in 
terms of the density $n$ and the scattering length $a$ of the gas as 
$(na^3)^{1/2}\ll 1$. Current gaseous condensates
satisfy such a condition. Nevertheless effects beyond the Gross-Pitaevskii
equation may be considered at zero temperature; also finite temperature
phenomena are not accounted for by the pure state mean field approach.

More complex theories have been developed in order to cope with effects
beyond the Gross-Pitaevskii equation:
Bogoliubov's approach takes into account the next term in the 
$(na^3)^{1/2}$ expansion \cite{bogol1,bogol2}. Also quantum kinetic theories have been developed to study
the formation of the condensate and to include the effect of the
non-condensed particles \cite{kinetic1,kinetic2,kinetic3}. Unfortunately the corresponding calculations 
are quite heavy for 3D non-homogeneous systems such as trapped gases,
and this constitutes a first limitation to the use of these methods. 
Also approximations used in some of these mean field theories are not
under rigorous control, making it difficult to assess their 
domain of validity (for a review see e.g.\  \cite{Dalfovo}).
Therefore a computational scheme
capable to provide exact results can have a great importance both from a 
purely theoretical point of view and for a quantitative analysis of
experimental data.

When the Bose gas is at thermal equilibrium 
such an exact numerical calculation of the properties
of the gas is available, using the Quantum 
Monte-Carlo techniques, based on
Feynman path integral formulation of quantum mechanics 
\cite{Krauth96,Ceperley}.
The aim of this paper is to present an alternative
exact and numerically tractable 
solution to the problem of the interacting Bose gas, a method
not restricted to the case of thermal equilibrium but which
allows for the study of the dynamics of the gas. 
The method is based on a stochastic evolution of Hartree states, in which 
all atoms have the same wave function, these Hartree states
being either Fock states (fixed number of atoms) or
coherent states. As a particular case of this solution with coherent states,
we recover the stochastic scheme corresponding to the evolution of the
density operator of the system in the positive $P$-representation 
\cite{Gardiner_livre,Scully,WallsMilburn,Steel}. 
This evolution is known to lead to strong unstability problems, which
fortunately do not show up for other implementations of the present method.

The outline of the paper is the following:
In section \ref{sec:stoch_form}, we present the stochastic formulation of the
evolution of these Hartree states which, 
after average over the stochastic component,
leads to the exact evolution. Section \ref{sec:Schemes} is devoted to the
presentation of two particular schemes 
implementing this stochastic formulation.
We first present a \lq simple' scheme, which minimizes the statistical spread of 
the calculated $N$-atom density matrix. 
We also investigate a more elaborate scheme
in which the trace of the calculated density matrix remains strictly 
constant in 
the evolution. With this constraint, we recover for coherent states 
the known stochastic simulation associated with the 
positive $P$-representation  
\cite{Steel}. 
Finally we investigate in sections \ref{sec:MC} and \ref{sec:1D} two examples, 
a two-mode model system and a one-dimensional Bose gas respectively. These
examples illustrate the accuracy and
the limitations of the method. Generally speaking we find that the \lq simple' scheme
simulations are only limited by the computation power: the number of realizations
needed for a good statistical accuracy increases exponentially with time for the simulation
with Fock states.
On the contrary the simulations with constant trace are subject to divergences
of the norms of the stochastic wave functions in finite time, a phenomenon
already known for coherent states in the context of the positive $P$-representation \cite{Christ}.

%%%%%%%%%%%%%%%%%%%%%%%%%%%%%%%%%%%%%%%%%%%%%%%%%%%%%%%%%%%%%%%%%%%%%

\section{Stochastic formulation of the 
$N$-boson problem using Hartree functions}\label{sec:stoch_form}

\subsection{Model considered in this paper}
The Hamiltonian of the trapped interacting Bose gas under exam can be 
written in terms of the Bose field operator 
$\Psih(x)$ as:
\begin{equation}\label{eq:Hamilt}
{\mathcal H}=\int\!dx\,\Psihd(x) h_0 \Psih(x)+\frac{1}{2}
\int\!\int\!dx\,dx'\,\Psihd(x)\Psihd(x')V(x-x')\Psih(x')\Psih(x)
\end{equation}
where $x$ is the set of spatial coordinates of a particle,
$h_0=-\frac{\hbar^2}{2m}\nabla^2+V_{\rm ext}(x)$ is the 
single particle Hamiltonian in the external confining potential 
$V_{\rm ext}$ and where
interactions are assumed to occur via a two-body potential $V(x-x')$.

In practice we consider the dilute gas and the low temperature regimes,
which correspond respectively to $n|a|^3\ll 1$ and
$|a|\ll \lambda$ for a three-dimensional
problem ($\lambda=h/(2\pi m k_B T)^{1/2}$
is the thermal de Broglie wavelength).
The true interaction potential can then be replaced by a simpler model 
potential leading to the same scattering length $a$ provided that
the range $b$ of this model potential is
much smaller than the healing length $\xi=(8\pi n a)^{-1/2}$ and than $\lambda$ \cite{bogol1,bogol2}.
This ensures that the physical results do not depend on $b$.
For simplicity we will use here repulsive
Gaussian potentials corresponding to a positive scattering length $a>0$.

\subsection{A stochastic Hartree Ansatz with Fock states}
 From a mathematical point of view, the exact evolution of the $N$-body 
density matrix $\rho$ can be obtained from the Hamiltonian \eq{Hamilt} 
using the quantum-mechanical equation of motion
\begin{equation}\label{eq:ExactEv}
\dot \rho(t)=\frac{1}{i\hbar}\left[{\mathcal H},\rho(t)\right]
\end{equation}
but any concrete calculation is impracticable even for moderate 
particle numbers $N$, due to the multi-mode nature of the problem
leading to a huge dimensionality of the $N-$body Hilbert space.

For this reason approximate theories have been developed in order to get 
useful results at least in some specific ranges of parameters;
the simplest one is the so-called mean-field theory, in which 
the $N$-particle density matrix is approximated by a Fock 
state Hartree ansatz
\begin{equation}\label{eq:HartreeAnsatz}
\rho(t)=\ketbra{N:\phi(t)}{N:\phi(t)}.
\end{equation}
The evolution of the normalized {\em condensate wave function} $\phi$ 
is determined using either a factorization approximation in the evolution
equation for the field operator \cite{bogol2} or a variational procedure \cite{CCT}.
The result is the well-known mean-field equation
\begin{equation}\label{eq:GP}
i\hbar\frac{\partial\phi(x)}{\partial t}=\left(-\frac{\hbar^2\nabla^2}{2m}+
V_{\rm ext}(x)\right)\phi(x)+(N-1)\left(\int\!dx'\;V(x-x')
\valass{\phi(x')}^2\right)\phi(x).
\end{equation}
For an interaction potential $V(x-x')$ 
modeled by a contact term $g\delta(x-x')$ (where $g=4\pi \hbar^2 a/m$ in a 
three-dimensional problem)
it reduces to the Gross-Pitaevskii equation
commonly used to analyze the dynamics of pure Bose-Einstein condensed gases.

A first attempt to improve the accuracy of the Hartree ansatz
\eq{HartreeAnsatz}
is to allow for a stochastic contribution $dB$ in the evolution 
of the macroscopic wave function $\phi$:
\begin{equation}\label{eq:HartreeEvStoch1}
\phi(t+d t)=\phi(t)+F\; d t + d B\quad.
\end{equation}
In all this paper the noise $dB$ is treated in the standard Ito formalism
\cite{Gardiner}: it is assumed to have a zero mean $\mean{dB}=0$ and to 
have a variance $\mean{dB^2}\propto dt$; 
a deterministic contribution is given by the \lq\lq force" term $Fdt$.
In this framework, the $N$-body density matrix would result 
from the stochastic mean over noise or, in other terms, 
from a mean over the probability distribution ${\mathcal P}(\phi)$ in the 
functional space of the wave functions $\phi$:
\begin{equation}\label{eq:HartreeAnsatzStoch1}
\rho(t) \stackrel{?}{=} 
\Big\langle{\ketbra{N:\phi(t)}{N:\phi(t)}}
\Big\rangle_{\rm stoch}
= \int\! {\mathcal D}\phi\;{\mathcal P}(\phi)\; \ketbra{N:\phi(t)}{N:\phi(t)}.
\end{equation}
An immediate advantage of this prescription over the pure
state ansatz Eq.(\ref{eq:HartreeAnsatz}) is that it could deal with
finite temperature problems \cite{Kagan}. However as shown in \S\ref{subsec:valid_stoch},
the simple generalization Eq.(\ref{eq:HartreeEvStoch1})
of the Gross-Pitaevskii equation cannot lead to an exact solution of the
$N$-body problem \cite{nonposit}.
Therefore we have to enlarge the family of dyadics over which we
expand the density operator; more precisely we use 
Hartree dyadics in which the wave functions in the bra and in the ket 
are different:
\begin{equation}\label{eq:DecorrAnsatz}
\sigma(t)=\ketbra{N:\phi_1(t)}{N:\phi_2(t)}\quad.
\end{equation}
The two wave functions $\phi_1(x)$ and $\phi_2(x)$ are assumed to 
evolve according to Ito stochastic differential equations:
\begin{equation}\label{eq:HartreeEvStoch2}
\phi_\alpha(t+d t)=\phi_\alpha(t)+F_\alpha\; d t + d B_\alpha 
\qquad (\alpha=1,2).
\end{equation}
The expansion Eq.(\ref{eq:HartreeAnsatzStoch1}) is then replaced by
\begin{equation}\label{eq:HartreeAnsatzStoch2}
\rho(t) = \Big\langle{
\ketbra{N:\phi_1(t)}{N:\phi_2(t)}}\Big\rangle_{\rm stoch}
= \int\!\!\int\! {\mathcal D}\phi_1\;{\mathcal D}\phi_2\;{\mathcal P}(\phi_1,\phi_2)\; 
\ketbra{N:\phi_1(t)}{N:\phi_2(t)}.
\end{equation}
We will see in the following that within this extended Hartree ansatz
one can find a stochastic evolution for $\phi_{1,2}$ reproducing
the exact time evolution. 

Actual calculations (see \S\ref{sec:MC} and \S\ref{sec:1D}) will be performed with a 
Monte-Carlo technique, in which the evolution of the probability 
distribution ${\mathcal P}$ is simulated by a large but finite number ${\mathcal N}$ of 
independent realizations $\phi_{1,2}\al{i}(t)$, $i=1,\ldots,{\cal N}$. At any time 
the (approximate) density matrix $\rho$ is given by the mean over such 
an ensemble of wave functions:
\begin{equation}\label{eq:StochMean}
\rho(t)\simeq \frac{1}{\mathcal N} \sum_{i=1}^{\mathcal N}  
\ketbra{N:\phi_1\al{i}(t)}{N:\phi_2\al{i}(t)}.
\end{equation}
The expectation value of any operator ${\hat O}$ is thus expressed by:
\begin{equation}\label{eq:StochMeanO}
\expect{\hat O}\simeq
\frac{1}{\mathcal N}\sum_{i=1}^{\mathcal N}
\braopket{N:\phi\al{i}_2(t)}{\hat O}{N:\phi\al{i}_1(t)}.
\end{equation}
For an Hermitian operator one can equivalently consider only the real part of this expression
since the imaginary part is vanishingly small in the large ${\cal N}$ limit.

Consider as an example
the one-particle density matrix of the gas, usually defined as:
\begin{equation}\label{eq:SingPartDensMat}
\rho\al{1}(x,x')=\expect{\Psihd(x')\Psih(x)}.
\end{equation}
Inserting in this expression our form of the complete density matrix 
\eq{StochMean}, we obtain the simple result
\begin{equation}\label{eq:StochSingPartDensMat}
\rho\al{1}(x,x')=N\;\Big\langle{
\phi_1(x)\phi_2^*(x')\braket{\phi_2}{\phi_1}^{N-1}}
\Big\rangle_{\rm stoch}
\end{equation}
from which it is easy to obtain the spatial density $n(x)=\rho\al{1}(x,x)$ 
and the correlation function 
$g\al{1}(x,x')=\rho\al{1}(x,x')/(n(x)n(x'))^{1/2}$.
Also, the condensate fraction can be 
obtained from the largest eigenvalue of $\rho\al{1}(x,x')$.

\noindent {\bf Remarks: }
\begin{enumerate}
\item The desired stochastic evolution, which has to satisfy Tr$[\rho]=1$,
cannot preserve the normalization of
$\phi_{1,2}$ to unity; we can write indeed
\begin{equation}
\mbox{Tr}[\rho(t)]=\Big\langle{
\langle\phi_2(t)|\phi_1(t)\rangle^N}\Big\rangle_{\rm stoch}=1
\end{equation}
which for $|\phi_1\rangle\neq |\phi_2\rangle$ imposes
$||\phi_1||\; ||\phi_2|| >1$.
\item The expansion Eq.(\ref{eq:HartreeAnsatzStoch2}) is always possible.
Using the identity
\begin{equation}
\mbox{Id}_N  = \lim_{{\cal M}\rightarrow +\infty}
\frac{1}{\cal M} \sum_{j=1}^{\cal M}|N:\psi^{(j)}\rangle\langle N:\psi^{(j)}|
\end{equation}
where the functions $\psi^{(j)}$ have a uniform distribution over the unit sphere
in the functional space,
we obtain
\begin{equation}
\rho = \lim_{{\cal M}\rightarrow +\infty} \frac{1}{{\cal M}^2}
\sum_{j_1,j_2=1}^{\cal M} |N:\psi^{(j_1)}\rangle\langle N:\psi^{(j_2)}|
\, \langle N:\psi^{(j_1)}|\rho|N:\psi^{(j_2)}\rangle.
\end{equation}
We write the matrix elements $\langle N:\psi^{(j_1)}|\rho|N:\psi^{(j_2)}\rangle$
as $\xi_{(j_1,j_2)}^{2N}$  and we set 
$\phi_1^{(j_1,j_2)} = \psi^{(j_1)} \xi_{(j_1,j_2)}$ 
and $\phi_2^{(j_1,j_2)}=\psi^{(j_2)} \xi^*_{(j_1,j_2)}$.  
Putting ${\cal N}={\cal M}^2$ and reindexing $(j_1,j_2)$ as a single
index $i$ we recover the expansion Eq.(\ref{eq:StochMean}).
Note that this expansion is not unique and does not have the pretension
to be the most efficient one.
For instance if the system is initially in a Hartree state $|N:\phi_0\rangle$,
such a procedure is clearly not needed since one has just to set 
$\phi_1^{(i)}(t=0)=\phi_2^{(i)}(t=0)=\phi_0$. This will be the case
of the numerical examples in sections \ref{sec:MC} and \ref{sec:1D}.
\end{enumerate}

%%%%%%%%%%%%%%%%%%%%%%%%%%%%%%%%%%%%%%%%%%%%%%%%%%%%%%%%%%%%%%%%

\subsection{Stochastic evolution of a Fock state Hartree dyadic}

In this subsection we calculate the stochastic time evolution during an 
infinitesimal time interval $dt$ of the dyadic $\sigma(t)$ 
given in Eq.(\ref{eq:DecorrAnsatz}). This will be used later in a comparison
with the exact master equation.

After $d t$, the dyadic $\sigma$ has evolved into: 
\begin{equation}\label{eq:Sigmat+dt}
\sigma(t+dt)=\ketbra{N:\phi_1 + d \phi_1}{N:\phi_2 +d \phi_2},
\end{equation}
where $d \phi_1$ and $d \phi_2$, defined according to 
\eq{HartreeEvStoch2}, 
contain both the deterministic contribution $F_\alpha dt$ and the 
stochastic one $dB_\alpha$.
Splitting each contribution into a longitudinal and an orthogonal component 
with respect to $\phi_\alpha$ and isolating a Gross-Pitaevskii term in the deterministic 
contribution, we can write:
\begin{eqnarray}
&&d B_\alpha(x)=\phi_\alpha(x)\,d\gamma_\alpha+d B_\alpha^\perp(x) \label{eq:StochdB} \\
&&F_\alpha(x)=F_\alpha^{GP}(x)+\lambda_\alpha\phi_\alpha(x)+F_\alpha^\perp(x).
\label{eq:DetF} 
\end{eqnarray}
Our choice of the Gross-Pitaevskii term is the following one:
\begin{eqnarray}
F_\alpha^{GP}(x)=&&\frac{1}{i\hbar}\left[h_0+\frac{(N-1)}{
\norme{\phi_\alpha}^2}\int\!dx'\;V(x-x')\valass{\phi_\alpha(x')}^2
\right]\phi_\alpha(x) \nonumber \\
&-&
\frac{1}{i\hbar}\left[\frac{(N-1)}{2}\frac{\braopket{\phi_\alpha
\phi_\alpha}{V}{\phi_\alpha\phi_\alpha}}{\norme{\phi_\alpha}^4}
\right]\phi_\alpha(x). \label{eq:GPcorr}
\end{eqnarray}
The first term gives the standard Gross-Pitaevskii evolution, including the kinetic term, 
the potential energy of the trap and the mean-field interaction energy; 
the second term, which arises naturally because we are considering
Fock states (rather than coherent states as commonly done)
takes into account the difference between the total mean-field energy 
per particle of the condensate and its chemical potential 
$\mu$~\cite{quelqu'un}.

We split the field operator in its longitudinal and transverse components,
keeping in mind that the wave functions $\phi_\alpha$ are not
of unit norm:
\begin{equation}\label{eq:DecompPsi}
\Psihd (x) = \frac{\phi_\alpha^*(x)}
{\norme{\phi_{\alpha}}^2}\hat{a}_{\phi_\alpha}^\dagger
+\delta \Psihd_\alpha(x)
\end{equation}
with 
\begin{equation}
\hat{a}_{\phi_\alpha}^\dagger = \int dx\;\phi_{\alpha}(x) \; \Psihd(x).
\end{equation}
The relevant bosonic commutation relations then read:
\begin{equation}
[\hat{a}_{\phi_\alpha},\hat{a}_{\phi_\alpha}^\dagger] = \norme{\phi_{\alpha}}^2
\qquad \mbox{and} \qquad [\hat{a}_{\phi_\alpha},\delta \Psihd_\alpha(x)]=0.
\end{equation}
We will also need the projector ${\mathcal Q}_{\alpha}$ 
onto the subspace orthogonal to
$\phi_{\alpha}$:
\begin{equation}
{\mathcal Q}_{\alpha}^{(x)}[ \psi(x,x',\ldots)]= \psi(x,x',\ldots)
-\frac{\phi_\alpha(x)}{||\phi_\alpha||^2}\int dy\;
\phi_\alpha^*(y)\,\psi(y,x',\ldots).
\end{equation}
This projector arises in the calculation as we have introduced
a component of the field operator orthogonal to $\phi_\alpha$.
Using $\int\!dx\; \phi_\alpha(x)\;\delta \Psihd_\alpha(x)=0$
we shall transform integrals
involving $\delta \Psihd_\alpha(x)$ as follows:
\begin{equation}
\int\!dx\; \psi(x,x',\ldots) \delta \Psihd_\alpha(x)=
\int\!dx\; {\mathcal Q}_{\alpha}^{(x)}[ \psi(x,x',\ldots)]\delta \Psihd_\alpha(x).
\end{equation}

Inserting these definitions in \eq{Sigmat+dt} 
the expression for $\sigma$ at time $t+dt$ can be written as
\begin{eqnarray}
\sigma(t+d t)-\sigma(t)=&&S\al{0}_1\ketbra{N:\phi_1}{N:\phi_2}+
\textrm {e.c.} \nonumber \\
&+&\int\!dx\;S\al{1}_1(x)\delta\Psihd_1(x)\ketbra{N-1:\phi_1}{N:\phi_2}+
\textrm{e.c.} \nonumber \\
&+&\int\!\!\int\!dx\;dx'\;S\al{2}_1(x,x')\delta\Psihd_1(x)\delta\Psihd_1(x')
\ketbra{N-2:\phi_1}{N:\phi_2}+\textrm{e.c.} \nonumber \\
&+&\int\!\!\int\!dx\;dx'\;S\al{1,1}(x,x')\delta
\Psihd_1(x)\ketbra{N-1:\phi_1}{N-1:\phi_2}\delta \Psih_2(x')
\label{eq:evstoch}
\end{eqnarray}
where the notation {\em e.c.} stands for the {\em exchanged} and 
{\em conjugate} of a quantity, that is
the complex conjugate of the same quantity after having 
exchanged the indices 1 and 2.
The explicit expressions for the 
$S\al{i}_\alpha$ are:
\begin{eqnarray}
&&S\al{0}_1=N
\frac{\braket{\phi_1}{F_1^{GP}}}{\norme{\phi_1}^2}
dt
+N\lambda_1\;dt+Nd\gamma_1+\frac{N(N-1)}{2}d 
\gamma_1^2+\frac{N^2}{2} d\gamma_1\;d\gamma_2^* \\
&&S\al{1}_1(x)=\sqrt{N}\left\{{\mathcal Q}_1^{(x)}\left[F_1^{GP}(x)\right]dt+
F_1^\perp(x)dt+
dB_1^\perp(x)+(N-1)d \gamma_1\; d B_1^\perp(x)
+N\;d B_1^\perp(x)\;d\gamma_2^*\right\} 
\label{toto} \\
&&S\al{2}_1(x,x')=\frac{\sqrt{N(N-1)}}{2}d B_1^\perp(x)\;d B_1^\perp(x') \\
&&S\al{1,1}(x,x')=N\;d B_1^\perp(x)\;d B_2^{\perp *}(x').
\label{eq:evstoch2}
\end{eqnarray}
Analogous expressions for $S_2^{(0)}$, $S_2^{(1)}$, $S_2^{(2)}$ are
obtained by exchanging the indices 1 and 2.
In the next subsection, we evaluate the exact evolution of the
same dyadic during a time interval $dt$, so that we can determine the 
constraints on the force and noise terms entering into these equations.

%%%%%%%%%%%%%%%%%%%%%%%%%%%%%%%%%%%%%%%%%%%%%%%%%%%%%%%%%

\subsection{Exact evolution of a Fock state Hartree dyadic}

To make the stochastic scheme described in the previous sections 
equivalent to the exact dynamics as it is given by \eq{Hamilt}, the 
final result of the previous subsection \eq{evstoch}-\eq{evstoch2} 
has to be compared with the exact evolution of the density matrix $\sigma(t)$.
Consider a dyadic $\sigma=\ketbra{N:\phi_1}{N:\phi_2}$ at time $t$; according 
to the equation of motion \eq{ExactEv}, 
after an infinitesimal time step $dt$ 
it has evolved into:
\begin{eqnarray}
\sigma(t+d t) &=& \sigma(t) + 
\frac{d t}{i\hbar} \left( {\cal H} \sigma(t) - \sigma(t) {\cal H}\right) \nonumber \\
&=&\sigma(t)+E\al{0}_1\ketbra{N:\phi_1}{N:\phi_2}+\textrm{e.c.} \nonumber \\
&+&\int\!dx\;E\al{1}_1(x)\delta\Psihd_1(x)
\ketbra{N-1:\phi_1}{N:\phi_2}+\textrm{e.c.} \nonumber \\
&+&\int\!\!\int\!dx\;dx'\;E\al{2}_1(x,x')\delta\Psihd_1(x)\delta\Psihd_1(x')
\ketbra{N-2:\phi_1}{N:\phi_2}+\textrm{e.c.}
\label{eq:evexacte}
\end{eqnarray}
where the $E\al{i}_\alpha$ are given by
\begin{eqnarray}
&&E\al{0}_1=\frac{N\,dt}{i\hbar}\left[
\frac{\braopket{\phi_1}{h_0}{\phi_1}}{\norme{\phi_1}^2}+
\frac{(N-1)}{2}
\frac{\braopket{\phi_1\phi_1}{V}{\phi_1\phi_1}}{\norme{\phi_1}^4}\right]\\
&&E\al{1}_1(x)=\frac{dt\sqrt{N}}{i\hbar}
{\mathcal Q}_1^{(x)}\left[
\left(h_0+ \frac{(N-1)}{\norme{\phi_1}^2}\;\int\!dx'\;V(x-x')
\valass{\phi_1(x')}^2\right)\phi_1(x)\right] \\
&&E\al{2}_1(x,x')=\frac{dt\sqrt{N(N-1)}}{2i\hbar}\;
{\mathcal Q}_1^{(x)}{\mathcal Q}_1^{(x')}[V(x-x')\phi_1(x)\phi_1(x')].
\label{eq:evexacte2}
\end{eqnarray}
Analogous expressions for $E_2^{(0)}$, $E_2^{(1)}$, $E_2^{(2)}$ are
obtained by exchanging the indices 1 and 2.

\subsection{Validity conditions for the stochastic Fock state Hartree ansatz}
\label{subsec:valid_stoch}

The similarity of the structures of \eq{evstoch} and \eq{evexacte} 
suggests the possibility of a stochastic scheme equivalent to the 
exact evolution: to achieve this, it is necessary to find out specific 
forms of deterministic \eq{DetF} and stochastic 
\eq{StochdB} terms for which the mean values of the $S_\alpha\al{i}$ 
equal the $E_\alpha\al{i}$:
\begin{eqnarray}
&&\mean{S\al{0}_1+S^{(0)*}_2}=E\al{0}_1+E^{(0)*}_2 \label{eq:condit1} \\
&&\mean{S\al{1}_\alpha(x)}=E\al{1}_\alpha(x) 
\label{eq:condit2} \\
&&\mean{S\al{2}_\alpha(x,x')}=
E\al{2}_\alpha(x,x') \label{eq:condit3} \\
&& \mean{S\al{1,1}(x,x')}=0 .
\label{eq:condit4}
\end{eqnarray}

 From the last equation \eq{condit4}, 
it follows immediately why independent bras and kets are needed in the 
ansatz \eq{DecorrAnsatz}: in the case $\phi_1=\phi_2=\phi$ such a 
condition would in fact lead to a vanishing orthogonal noise and 
finally to the impossibility of satisfying \eq{condit3}.

In terms of the different components, these conditions can be rewritten as:
\begin{eqnarray}
&&\left(\lambda_1+\lambda_2^*\right)d t+\frac{(N-1)}{2}\left[\mean{d 
\gamma_1^2}+\mean{d \gamma_2^2}^*\right]
+N\overline{d\gamma_1 d\gamma_2^*} =0 \label{eq:cond1} \\
&&F_1^\perp(x)\;d t+(N-1)\mean{d B_1^\perp(x)\,d\gamma_1}
+ N \mean{d B_1^\perp(x)\,d\gamma_2^*} =0 \label{eq:cond2} \\
&&F_2^\perp(x)\;d t+(N-1)\mean{d B_2^\perp(x)\,d\gamma_2}
+ N \mean{d B_2^\perp(x)\,d\gamma_1^*} =0 \label{eq:cond2bis} \\
&&\mean{d B_\alpha^\perp(x)d B_\alpha^\perp(x')}=\frac{d t}{i\hbar}
{\mathcal Q}_\alpha^{x} {\mathcal Q}_\alpha^{x'}
\left[V(x-x') \phi_\alpha(x)\phi_\alpha(x')\right] \label{eq:cond3} \\
&&\mean{d B_1^\perp(x) {d B_2^\perp}^*(x')}=0.
\label{eq:cond4}
\end{eqnarray}
As we shall discuss in detail in \S\ref{sec:Schemes}, 
several different stochastic schemes can be found satisfying
 \eq{cond1}-\eq{cond4}; each of them gives an evolution 
identical in average to the exact one, but the 
statistical properties can be very different.

%%%%%%%%%%%%%%%%%%%%%%%%%%%%%%%%%%%%%%%%%%%%%%%%%%%%%%%%

\subsection{A stochastic Hartree ansatz with coherent states}

Up to now we have worked out the case of a Fock state 
ansatz $\ketbra{N:\phi_1}{N:\phi_2}$.
Actually coherent states rather than Fock states are generally
used, both in quantum optics and in condensed matter physics.
We now show that our stochastic procedure also applies
with a coherent state ansatz of the form:
\begin{equation}\label{eq:CoherAnsatz}
\sigma(t)=\Pi(t)\;\ketbra{\textrm{ coh}: \phi_1}{\textrm{ coh}: \phi_2},
\end{equation}
with
\begin{equation}\label{eq:cohstate}
\ket{\textrm{ coh}: \phi_\alpha}=
\exp\left(\bar{N}^{1/2}\int\! dx\;\phi_\alpha(x)\Psihd(x)\right)\ket{{\rm vac}}
\end{equation}
where $\bar{N}$ is the mean number of particles.
We have included here a prefactor $\Pi(t)$ which was absent in the
case of the Fock state ansatz Eq.\eq{DecorrAnsatz}; in the Fock state case 
indeed such a prefactor could be reincluded into the definition
of $\phi_1$ and $\phi_2$.
The wave functions $\phi_\alpha(x)$ 
and the prefactor factor $\Pi$ evolve according to Ito stochastic 
differential equations
\begin{eqnarray}
d \phi_\alpha=F_\alpha d t+d B_\alpha \nonumber \\
d \Pi=f\,d t+d b.
\label{eq:cohevolst}
\end{eqnarray}
Splitting the field operator as
\begin{equation}
\Psih(x)=\bar{N}^{1/2}\; \phi_\alpha(x) + \delta\Psih_\alpha(x)
\end{equation}
and using 
\begin{equation}
\delta\Psih_\alpha(x)\ket{\textrm{ coh}: \phi_\alpha} = 0
\end{equation}
we find that
the equivalence of the stochastic scheme and the exact evolution 
translates into the set of conditions:
\begin{eqnarray}
&&f=0 \label{eq:condcoh1} \\
&&F_1(x)d t+\frac{1}{\Pi}\mean{d b\, d B_1(x)}
=\frac{d t}{i\hbar}h_0\phi_1(x) \label{eq:condcoh2} \\
&&F_2(x) d t+\frac{1}{\Pi^*}\mean{d b^*\, d B_2(x)}
=\frac{d t}{i\hbar}h_0\phi_2(x)  \label{eq:condcoh2bis} \\
&&\mean{d B_\alpha(x) d B_\alpha(x')}
=\frac{d t}{i\hbar}V(x-x')\phi_\alpha(x)\phi_\alpha(x') \label{eq:condcoh3} \\
&&\mean{d B_1(x)d B_2^*(x')}=0. 
\label{eq:condcoh4}
\end{eqnarray}
As we shall see in \S\ref{sec:Schemes}, such conditions are 
satisfied by several stochastic schemes. 
Very remarkably, the stochastic evolution deduced from the positive $P$-representation
\cite{PositiveP} arises naturally as one of them.

Within this coherent state ansatz 
the one-particle density matrix is evaluated using
\begin{equation}\label{eq:OnePartCoh}
\rho\al{1}(x,x')=
\bar{N} \Big\langle 
\phi_1(x)\phi_2^*(x')\;\Pi(t)
\;\exp(\bar{N}\braket{\phi_2}{\phi_1})\Big\rangle_{\rm stoch}.
\end{equation}
In a practical implementation of the simulation it turns out to be numerically
more efficient to represent $\Pi(t)$ as the exponential of some quantity
\begin{equation}
\Pi(t)=e^{\bar{N} S(t)}
\end{equation}
and to evolve $S(t)$ according to the stochastic equation
\begin{eqnarray}
d S &=& -\int dx\; [dB_1(x)\phi_1^*(x)+dB_2^*(x)\phi_2(x)]\nonumber \\
&&-\frac{\bar{N}dt}{2i\hbar}\int dx\int dx'\; V(x-x') 
\left[|\phi_1(x)|^2|\phi_1(x')|^2-|\phi_2(x)|^2|\phi_2(x')|^2 \right].
\end{eqnarray}

%%%%%%%%%%%%%%%%%%%%%%%%%%%%%%%%%%%%%%%%%%%%%%%%%%%%%%%%%%%%%%%%
%%%%%%%%%%%%%%%%%%%%%%%%%%%%%%%%%%%%%%%%%%%%%%%%%%%%%%%%%%%%%%%%

\section{Particular implementations of the stochastic approach}
\label{sec:Schemes}

In the previous section we have derived the conditions that a stochastic 
scheme has to satisfy in order to recover the exact evolution given by the 
Hamiltonian \eq{Hamilt}; in the case of the Fock state 
ansatz \eq{DecorrAnsatz}, we get to the system \eq{cond1}-\eq{cond4}, 
while in the case of the coherent state ansatz \eq{CoherAnsatz} 
we get to the conditions \eq{condcoh1}-\eq{condcoh4}.
As the number of these equations is actually smaller than the number
of unknown functions there is by no mean uniqueness of the solutions, that
is of the simulation schemes. We need a strategy to identify interesting
solutions.

We therefore 
start this section by considering various indicators of the statistical error
of the simulation (\S\ref{subsec:growth}) which can be used as guidelines
in the search for simulation schemes.
These indicators are defined as variances of
relevant quantities which are conserved in the exact evolution but which
may fluctuate in the simulation. We show that these indicators are non
decreasing functions of time; attempts to minimize the time derivative
of a specific indicator will lead to particular implementations of
the general stochastic method, 
such as the {\em simple} scheme (\S\ref{subsec:simple})
and the {\em constant trace} scheme (\S\ref{subsec:constant}).

\subsection{Growth of the statistical errors} \label{subsec:growth}

The first indicator that we consider measures the squared deviation of 
the stochastic operator $\sigma(t)$ from the exact density operator $\rho(t)$:
\begin{eqnarray}
\Delta(t) &=& \Big\langle {\mbox{Tr}[(\sigma^{\dagger}(t)-\rho(t))
(\sigma(t)-\rho(t))]}\Big\rangle_{\rm stoch}  \nonumber \\
&=& \Big\langle{\mbox{Tr}[\sigma^{\dagger}(t)\sigma(t)]}\Big\rangle_{\rm stoch}
- \mbox{Tr}[\rho(t)^2].
\label{eq:def_Delta}
\end{eqnarray}

We now show that $\Delta(t)$
is a non-decreasing function of time.
When the stochastic scheme satisfies the validity conditions derived
in the previous section,
we can write the stochastic equation for $\sigma$ as:
\begin{equation}\label{eq:evol_compacte}
d\sigma = \frac{dt}{i\hbar}[{\mathcal H},\sigma] + d\sigma_s
\end{equation}
where $d\sigma_s$ is a  zero-mean noise term linear in $dB_\alpha$ (and $db$ 
for the coherent state simulation). In the case of simulation with Fock states
it is given by
\begin{equation}
d\sigma_s = N^{1/2} \left\{
\int dx\; dB_1(x) \hat{\Psi}^{\dagger}(x) |N-1:\phi_1\rangle
\langle N:\phi_2| + \int dx\; dB_2^*(x) |N:\phi_1\rangle \langle N-1:\phi_2|
\hat{\Psi}(x)\right\}.
\label{eq:dssf}
\end{equation}
In the case of simulation with coherent states it is given by
\begin{eqnarray}
d\sigma_s &=& db |{\rm coh}\;:\phi_1\rangle\langle {\rm coh}\;:\phi_2| \nonumber\\
&+& \bar{N}^{1/2}\Pi\left\{
\int dx\; dB_1(x) \hat{\Psi}^{\dagger}(x) 
|{\rm coh}\;:\phi_1\rangle\langle {\rm coh}\;:\phi_2|
+\int dx\; dB_2^*(x)
|{\rm coh}\;:\phi_1\rangle\langle {\rm coh}\;:\phi_2|
\hat{\Psi}(x) \right\}.
\label{eq:dssc}
\end{eqnarray}

We calculate the variation of $\Delta$ during $dt$, 
replacing $\sigma$ by $\sigma+d\sigma$
in Eq.(\ref{eq:def_Delta}) and keeping terms up to order $dt$. Using the invariance of
the trace in a cyclic permutation and averaging over the noise between
$t$ and $t+dt$ we finally obtain
\begin{equation}
\label{eq:dDelta}
d\Delta= 
\Big\langle{\mbox{Tr}[d\sigma^{\dagger}_sd\sigma_s]}\Big\rangle_{\rm stoch},
\end{equation}
which is a positive quantity. Minimization of this quantity
is the subject of \S\ref{subsec:simple}.
Physically $d\Delta\geq 0$  means that the 
impurity of the stochastic density operator
$\sigma$ always increases in average, while the exact density operator has a constant
purity $\mbox{Tr}[\rho^2]$.

The second kind of indicator that we consider measures the statistical error
on constants of motion of the exact evolution. Consider a time independent operator $X$
commuting with the Hamiltonian.
The stochastic evolution leads to an error on the expectation value of $X$
with a variance given by the ensemble average of
\begin{eqnarray}
\Delta_X(t) &=& \Big\langle{\Big|\mbox{Tr}[X\sigma(t)]-\mbox{Tr}[X\rho(t)]\Big|^2}
\Big\rangle_{\rm stoch}\nonumber
\\
&=& \Big\langle{\Big|\mbox{Tr}[X\sigma(t)]\Big|^2}\Big\rangle_{\rm stoch} 
- \Big|\mbox{Tr}[X\rho(t)]\Big|^2.
\end{eqnarray}
 From Eq.(\ref{eq:evol_compacte}) we obtain the variation after a time step $dt$
of the expectation value of $X$ along a stochastic trajectory:
\begin{equation}\label{eq:d_tr}
d(\mbox{Tr}[X\sigma]) =  \frac{dt}{i\hbar}\mbox{Tr}\left(X[{\mathcal H},\sigma]\right)
+\mbox{Tr}[Xd\sigma_s].
\end{equation}
Using the invariance of the trace under cyclic permutation and the commutation
relation $[{\mathcal H},X]=0$ we find that the first term in the right hand side
of Eq.(\ref{eq:d_tr}) vanishes so that
\begin{equation}
d\Delta_X =  \Big\langle\Big|\mbox{Tr}[Xd\sigma_s]\Big|^2\Big\rangle_{\rm stoch},
\label{eq:expres}
\end{equation}
a quantity which is always non-negative.

Using expression (\ref{eq:expres}) one can \lq design' simulations preserving exactly 
the conserved quantity, the constraint to meet being $\mbox{Tr}[Xd\sigma_s]=0$:
for instance in the Fock state simulation,
one first chooses the transverse noises $dB_\alpha^\perp$ satisfying
Eqs.(\ref{eq:cond3},\ref{eq:cond4}); then one simply has
to take for the longitudinal noise of $\phi_1$:
\begin{equation}
d\gamma_1 = -\frac{1}{\sqrt{N}}(\langle N:\phi_2|X|N:\phi_1\rangle)^{-1}
\;\int dx\; dB_1^\perp(x) \langle N:\phi_2|X\delta\hat{\Psi}_1^\dagger(x)|N-1:\phi_1\rangle
\label{eq:dgX}
\end{equation}
and a similar expression for $d\gamma_2$; finally the force terms $F_\alpha$
are adjusted in order to satisfy Eq.(\ref{eq:cond1}-\ref{eq:cond2bis}).
As natural examples of conserved quantities one can choose
$X=1$ or $X={\cal H}$; the former case is discussed in detail in \S\ref{subsec:constant}.

\subsection{The {\em simple} schemes} \label{subsec:simple}
These schemes are characterized by the minimization of the incremental variation
of the statistic spread of the $N$-particle density matrix $\sigma(t)$, a
spread that we have
already quantified in Eq.(\ref{eq:def_Delta}) by $\Delta(t)$. To be more specific
we assume that we have evolved a dyadic up to time $t$, and we look for the noise
terms that minimize the increase of $\mbox{Tr}[\sigma^\dagger\sigma]$ 
between $t$ and $t+dt$.

\subsubsection{Simulation with Fock states}

In the case of the Fock state ansatz, we calculate explicitly the variation
of $\mbox{Tr}[\sigma^\dagger\sigma]$
from Eq.(\ref{eq:dssf}) and we get:
\begin{equation}\label{eq:dDF}
\frac{d\mbox{Tr}[\sigma^\dagger\sigma]}{N\mbox{Tr}[\sigma^\dagger\sigma]}=
N\overline{|d\gamma_1+d\gamma_2^*|^2}
+\sum_{\alpha=1,2}||\phi_\alpha||^{-2}
\int dx\; \overline{|dB_\alpha^\perp(x)|^2}
+\left[d\gamma_1+d\gamma_2^*+\mbox{c.c.}\right].
\end{equation}
We now look for the noise terms $d\gamma_\alpha$ and $dB_\alpha^\perp$ minimizing
this quantity under the constraints Eqs.(\ref{eq:cond1}-\ref{eq:cond4}).

We first note that we can choose $d\gamma_1=d\gamma_2=0$ without affecting
the transverse noises, as shown by Eqs.(\ref{eq:cond1}-\ref{eq:cond4}): the
correlation function of the transverse noises do not involve the $d\gamma_\alpha$,
and we can accommodate for any choice of $d\gamma_\alpha$ by defining appropriately
the force terms $F_\alpha^\perp,\lambda_\alpha$. In the particular case defining
our simple scheme we take all these force terms equal to zero.
Note that the choice of vanishing $d\gamma$'s immediately leads to
a vanishing noise term in Eq.(\ref{eq:dDF}).

Secondly the terms involving the transverse noise in Eq.(\ref{eq:dDF})
are bounded from below:
As the modulus of a mean is less than the mean of the modulus, we have
\begin{equation}
\label{eq:inegalite}
\left|\;\overline{dB_\alpha^\perp(x)dB_\alpha^\perp(x)}\;\right|\leq
\overline{\;|dB_\alpha^\perp(x)|^2\;},
\end{equation}
with the left hand side of this inequality fully determined
by condition Eq.(\ref{eq:cond3}).

We have found for the transverse noise a choice which fulfills Eqs.(\ref{eq:cond3},\ref{eq:cond4}) 
and which saturates the inequality
Eq.(\ref{eq:inegalite}):
\begin{equation}\label{eq:simple}
dB_\alpha^\perp(x)=\left(\frac{d t}{i\hbar}\right)^{1/2}
{\mathcal Q}_\alpha^{(x)} \left[\phi_\alpha(x)\,\int\!\frac{dk}{(2\pi)^{d/2}}
\left(\tilde{V}(k)\right)^{1/2}e^{ikx}e^{i\theta_\alpha(k)}\right]
\end{equation}
where $d$ is the dimension of position space,
$\tilde{V}(k)$ is the Fourier transform of the model interaction potential, supposed
here to be positive:
\begin{equation}
\tilde{V}(k) = \int dx\; V(x) e^{-ikx}.
\end{equation}
The phases $\theta_\alpha$ have the following statistical property: 
\begin{equation}\label{eq:theta}
\mean{e^{i\theta_\alpha(k)}e^{i\theta_\alpha(k')}}=\delta(k+k')
\end{equation}
and $\theta_1,\theta_2$ are uncorrelated.
In practice for half of the $k$-space (e.g.\ $k_z>0$) 
$\theta_\alpha(k)$ is randomly chosen between $0$ and $2\pi$;
for the remaining $k$'s we take $\theta_\alpha(-k)=-\theta_\alpha(k)$.
One can then check that this choice for the transverse noise reproduces
the correlation function Eqs.(\ref{eq:cond3},\ref{eq:cond4}). 

We show now that the implementation (\ref{eq:simple}) saturates the inequality 
Eq.(\ref{eq:inegalite}), so that it leads to the minimal possible value
for $d\mbox{Tr}[\sigma^\dagger\sigma]$
within the validity constraints of the stochastic approach.
We calculate explicitly the right hand side
of Eq.(\ref{eq:inegalite}):
\begin{eqnarray}
\overline{|dB_\alpha^{\perp}(x)|^2} &=&
\frac{dt}{\hbar}
\left({\mathcal Q}_\alpha^{(x)*} {\mathcal Q}_\alpha^{(x')} 
[V(x-x')\phi_\alpha^*(x)\phi_\alpha(x')]\right)_{x=x'} \nonumber \\
&=& \frac{dt}{\hbar} |\phi_\alpha(x)|^2 \left[
V(0) -2 \int dy\; \frac{|\phi_\alpha(y)|^2}{||\phi_\alpha||^2}V(x-y)
+\frac{\langle\phi_\alpha,\phi_\alpha|V|\phi_\alpha,\phi_\alpha\rangle}
{||\phi_\alpha||^4}\right]
\label{eq:rhs}
\end{eqnarray}
where ${\mathcal Q}_\alpha^{*}$ projects onto the subspace orthogonal to
$\phi_\alpha^*$ and where we have used the positivity of the Fourier
transform $\tilde{V}$ of the model interaction potential.
The left hand side of Eq.(\ref{eq:inegalite}) is calculated using
Eq.(\ref{eq:cond3}):
\begin{equation}
\overline{dB_\alpha^{\perp}(x)^2} = \frac{dt}{i\hbar} \phi_\alpha^2(x) \left[
V(0) -2 \int dy\; \frac{|\phi_\alpha(y)|^2}{||\phi_\alpha||^2}V(x-y)  
+\frac{\langle\phi_\alpha,\phi_\alpha|V|\phi_\alpha,\phi_\alpha\rangle}
{||\phi_\alpha||^4}\right].
\label{eq:lhs}
\end{equation}
As the expressions between square brackets in Eqs.(\ref{eq:lhs},\ref{eq:rhs})
are real positive we deduce the equality in Eq.(\ref{eq:inegalite}).

We can now calculate explicitly the variation of $\mbox{Tr}[\sigma^\dagger\sigma]$ 
by integrating Eq.(\ref{eq:rhs}) over $x$:
\begin{equation}
\frac{d\mbox{Tr}[\sigma^\dagger\sigma]}{N\mbox{Tr}[\sigma^\dagger\sigma]}
=  \frac{dt}{\hbar}\left[
2 V(0)-\sum_{\alpha=1,2}
\frac{\langle\phi_\alpha,\phi_\alpha|V|\phi_\alpha,\phi_\alpha\rangle}
{||\phi_\alpha||^{4}}\right].
\label{eq:dtrSF}
\end{equation}
This expression is particularly useful since it allows
one to derive an upper bound on the increase of $\mbox{Tr}[\sigma^\dagger\sigma]$:
as we assume here a positive Gaussian model potential $V(x-x')$
the matrix element $\langle\phi_\alpha,\phi_\alpha|V|\phi_\alpha,\phi_\alpha\rangle$ 
is positive so that the right hand side of Eq.(\ref{eq:dtrSF}) is smaller
than $2 V(0) dt/\hbar$. After time integration,
using Eq.(\ref{eq:def_Delta}) and the fact that the trace of the squared
density operator $\rho^2$ is a constant under Hamiltonian evolution we
can deduce an upper bound on the squared statistical error $\Delta(t)$:
\begin{equation}
\Delta(t) + \mbox{Tr}[\rho^2] \leq \left[\Delta(0)+\mbox{Tr}[\rho^2]\right]
e^{2NV(0)t/\hbar}.
\label{eq:maj_Delta}
\end{equation}
Note that it involves the model dependent quantity $V(0)$ and not only
the physical parameters of the problem such as the chemical potential
or the scattering length. It may be therefore important to adjust
the model interaction potential $V(x-x')$ in order to minimize the growth
of the statistical error for given physical parameters.

To summarize the proposed simple scheme has several noticeable
properties. The deterministic force acting on the $\phi_\alpha$'s is
simply the mean field contribution, so that the whole correction to the
mean field evolution is provided by the transverse noises $dB_\alpha^\perp$.
Also the evolutions of the two states $\phi_\alpha$ are totally independent
from each other.

%%%%%%%%%%%%%%%
\subsubsection{Simulation with coherent states}

In the case of the coherent state ansatz, an explicit calculation of 
$d\mbox{Tr}[\sigma^\dagger\sigma]$
from Eq.(\ref{eq:dssc}) gives:
\begin{eqnarray}
\frac{d\mbox{Tr}[\sigma^\dagger\sigma]}{\mbox{Tr}[\sigma^\dagger\sigma] } &=&
\overline{
\left|\frac{db}{\Pi}+\bar{N}\int dx\; dB_1(x)\phi_1^*(x)+dB_2^*(x)\phi_2(x)\right|^2}
+\bar{N}\sum_{\alpha=1,2}\int dx\;\overline{|dB_\alpha(x)|^2} \nonumber \\
&&+\left[\frac{db}{\Pi}+\bar{N}\int dx\; dB_1(x)\phi_1^*(x)+dB_2^*(x)\phi_2(x)
+\mbox{c.c}\right].
\label{eq:dDc}
\end{eqnarray}

We now proceed to the minimization of the increment of 
$\mbox{Tr}[\sigma^\dagger\sigma]$
within the coherent state ansatz along the same lines as the previous subsection.
First we optimize the noise $db$ on the normalization factor $\Pi$:
\begin{equation}
db=-\bar{N}\Pi\left(\int dx\; dB_1(x)\phi_1^*(x)+dB_2^*(x)\phi_2(x)\right).
\label{eq:condcohminbr3}
\end{equation}
This choice leads to a vanishing noise term 
in Eq.(\ref{eq:dDc}).
We insert this expression for $db$
in the validity conditions Eq.(\ref{eq:condcoh2})
and Eq.(\ref{eq:condcoh2bis}) and we obtain:
\begin{equation}
F_\alpha(x)=\frac{1}{i\hbar}\left[h_0+\bar{N}\int\!dx'\,V(x-x')
|\phi_\alpha(x')|^2\right]\phi_\alpha(x) \label{eq:condcohminbr1}.
\end{equation}

Finally minimization of the contribution of the noise terms $dB_\alpha$ with the
constraint Eq.(\ref{eq:condcoh3}) is achieved with the choice
\begin{equation}
dB_\alpha(x)=\left(\frac{d t}{i\hbar}\right)^{1/2}
\phi_\alpha(x)\,\int\!\frac{dk}{(2\pi)^{d/2}}
\left(\tilde{V}(k)\right)^{1/2}e^{ikx}e^{i\theta_\alpha(k)}
\label{eq:condcohminbr2} 
\end{equation}
where the phases $\theta_\alpha(k)$ are randomly generated as in Eq.(\ref{eq:theta}).

The first equation Eq.(\ref{eq:condcohminbr1})
fixes the deterministic evolution to the usual 
mean-field equation \eq{GP}. We note here
that the mean-field term in Eq.(\ref{eq:condcohminbr1}) does not contain the normalization
of the spatial density $\bar{N}|\phi_\alpha(x')|^2$ by $||\phi_\alpha||^2$, a feature
present in the Fock state simulation (see Eq.(\ref{eq:GPcorr})).
This is a disadvantage of the coherent state simulation since
this normalization factor appearing in the Fock state simulation
has a regularizing effect:
the norms  $||\phi_\alpha||$ may indeed deviate significantly from unity in the stochastic
evolution.
The second equation Eq.(\ref{eq:condcohminbr2})
determines the stochastic noise 
on the wave functions in a way very similar to the Fock state case
Eq.(\ref{eq:simple}). In particular the evolutions of $\phi_1$ and 
$\phi_2$ are still uncorrelated. The only difference is the disappearance of
the projector ${\mathcal Q}_\alpha$ in the expression of the noise, 
which leads to an increased
noise with respect to the simulation with Fock states.

As in the previous subsection we now estimate 
the squared error $\Delta$. 
We calculate the variation of $\mbox{Tr}[\sigma^\dagger\sigma]$
for the choice of noise Eq.(\ref{eq:condcohminbr2}):
\begin{equation}
\frac{1}{\mbox{Tr}[\sigma^\dagger\sigma]}\frac{d\mbox{Tr}[\sigma^\dagger\sigma]}{dt} = 
\frac{\bar{N} V(0)}{\hbar}
\sum_{\alpha=1,2} ||\phi_\alpha||^2.
\end{equation}
The average over all stochastic realizations of the norm squared
of the wave functions can be calculated exactly:
\begin{equation}
\Big\langle||\phi_\alpha||^2\Big\rangle_{\rm stoch}(t)=
e^{tV(0)/\hbar}
\Big\langle||\phi_\alpha||^2\Big\rangle_{\rm stoch}(0).
\end{equation}
This leads to a remarkable identity on the trace of $\sigma^\dagger\sigma$:
\begin{equation}
\Big\langle \ln \mbox{Tr}[\sigma^\dagger\sigma]\Big\rangle_{\rm stoch}(t)=
\Big\langle \ln \mbox{Tr}[\sigma^\dagger\sigma]\Big\rangle_{\rm stoch}(0)+
\bar{N}\left(e^{tV(0)/\hbar}-1\right)
\Big\langle \sum_{\alpha =1,2} ||\phi_\alpha||^2 \Big\rangle_{\rm stoch} (0).
\label{eq:lnSC}
\end{equation}
Using finally the concavity of the logarithmic function, leading to 
the logarithm of a mean being larger than the mean of the logarithm 
we obtain a lower bound on the squared error $\Delta$ on the $N-$body
density matrix:
\begin{equation}
\Delta(t) + \mbox{Tr}[\rho^2]\geq 
A \exp\left[2B\bar{N}\left(e^{tV(0)/\hbar}-1\right) \right]
\label{eq:min_Delta}
\end{equation}
where we have introduced the constant quantities
\begin{eqnarray}
A &=&
\exp\left[\Big\langle \ln \mbox{Tr}[\sigma^\dagger\sigma]\Big\rangle_{\rm stoch}(0)\right]
 \\
B &=& \frac{1}{2} \Big\langle \sum_{\alpha =1,2} ||\phi_\alpha||^2 \Big\rangle_{\rm stoch} (0)
\end{eqnarray}
It is quite remarkable that 
in the limit of times shorter than $\hbar/V(0)$ this lower bound scales
exponentially with time as the upper bound 
Eq.(\ref{eq:maj_Delta}) obtained for Fock states.
Consequently the simulation scheme with Fock states
is likely to be more efficient that the simulation with coherent states.
This will indeed be the case in the numerical examples given in the next sections.

\subsection{The {\em constant trace} schemes} \label{subsec:constant}

We have given the expression of the one-body density
matrix $\rho^{(1)}$ in terms of $\phi_\alpha(x)$ for the
simulation with Fock states Eq.(\ref{eq:StochSingPartDensMat}).
This expressions shows that $\rho^{(1)}$ is very sensitive 
in the large $N$ limit to fluctuations of $\langle\phi_2|\phi_1\rangle$. 
The same remark applies to two-body observables.
In order to improve the statistical properties of the simulation
one can consider the possibility of a simulation scheme
with $\langle\phi_2|\phi_1\rangle=1$ at any time.
This actually corresponds to a conserved trace of each single dyadic
$\sigma(t)$.
This possibility is analyzed in \S\ref{subsubsec:TCF}; it is
extended to the coherent state simulation in \S\ref{subsubsec:TCC},
leading to the well-known positive $P$-representation formalism.

\subsubsection{Simulation with Fock states} \label{subsubsec:TCF}
Within the Fock state ansatz, the conservation of 
the trace of the dyadic ${\rm Tr}[{\sigma}]$  can be achieved by (i) choosing
the transverse noises $dB_\alpha^\perp$ according to the formula Eq.(\ref{eq:simple})
and (ii) using the expression Eq.(\ref{eq:dgX}) for the longitudinal
noise with $X=1$. 
%Assuming that the density operator
%at time $t=0$ is expanded on dyadics with $\langle\phi_2|\phi_1\rangle=1$
Point (ii) gives:
\begin{equation}
d \gamma_1=-{\braket{\phi_{2}}{\phi_1}}^{-1}\;\int dx\;\phi_2^*(x)d B_1^\perp(x).
\label{eq:gamma1}
\end{equation}
The forces terms
$\lambda_\alpha$ and $F_\alpha$ are fixed by the conditions \eq{cond1}-\eq{cond2bis}:
\begin{equation}
\lambda_1 \;dt=
-\frac{N-1}{2}{\braket{\phi_{2}}{\phi_1}}^{-2}
\;\int dx\;dx'\;\phi_2^*(x)\phi_2^*(x')\mean{d B_1^\perp(x) d B_1^\perp(x')} 
\label{eq:lambda1}
\end{equation}
and
\begin{equation}\label{Fperp}
F_1^\perp(x)\;dt=
(N-1){\braket{\phi_{2}}{\phi_1}}^{-1}\;
\int dx'\;\phi_2^*(x')\mean{d B_1^\perp(x') d B_1^\perp(x)}.
\end{equation}
The expressions for $d\gamma_2, \lambda_2$ and $F_2^\perp(x)$ are obtained by exchanging 
the indices 1 and 2 in these results.

\subsubsection{Simulation with coherent states} \label{subsubsec:TCC}

In the case of the coherent state ansatz, the value of $d\sigma_s$,
which is the zero-mean noise term entering the variation of the
dyadic $\sigma$ during a time step $dt$, is given in
Eq.(\ref{eq:dssc}).
The requirement of a constant
trace ${\rm Tr}[{\sigma}]=\Pi e^{\bar{N}\braket{\phi_2}{\phi_1}}$  leads to
the following condition on the noise terms
\begin{equation}\label{eq:condcohconstr}
d b+\bar{N}\Pi\int dx\; \left(\phi_2^*(x)d B_1(x)+d B_2^*(x)\phi_1(x)\right)=0.
\end{equation}
We choose the noise terms $dB_\alpha$ as in Eq.(\ref{eq:condcohminbr2}).
The remaining parameters $F_\alpha$ are now unambiguously determined by
\eq{condcoh2}-\eq{condcoh2bis}:
\begin{eqnarray}
&&F_1(x)=\frac{1}{i\hbar}\left[h_0+\bar{N}\int\!dx'\,
\phi_2^*(x')V(x-x')\phi_1(x')\right]\phi_1(x) \label{eq:F1PP} \\
&&F_2(x)=\frac{1}{i\hbar}\left[h_0+\bar{N}\int\!dx'\,
\phi_1^*(x')V(x-x')\phi_2(x')\right]\phi_2(x).
\label{eq:F2PP}
\end{eqnarray}
This scheme exactly recovers the stochastic evolution
in the positive $P$-representation, 
which was originally obtained with a different mathematical procedure~\cite{PositiveP}.

\section{Stochastic vs. exact approach for a two-mode model}
\label{sec:MC}

In order to test the convergence of the stochastic schemes 
developed in the previous section we now apply this method to
a simple two-mode system for which the exact solution of the
$N$-body Schr\"odinger equation can also be obtained by a direct numerical
integration.
This allows (i) to check that the stochastic methods when averaged over
many realizations give the correct result indeed, and (ii) to determine
the statistical error for each of the four implementations of
the stochastic approach (\lq constant trace' vs.\ \lq simple', Fock vs.\ coherent
states).

The toy-model that we consider describes the dynamics of two self-interacting
condensates coherently coupled one to the other. It can be applied to
the case of two condensates separated by a barrier 
\cite{Kasevich} (Josephson-type
coupling) or condensates in two different internal states coupled
by an electromagnetic field \cite{JILA_Rabi} (Rabi-type coupling).
In this model we restrict the expansion of the atomic field operator to
two orthogonal modes,
\begin{equation}
\hat{\psi}(x) = \hat{a}\, u_a(x) + \hat{b}\, u_b(x).
\end{equation}
The Hamiltonian Eq.(\ref{eq:Hamilt}) takes the simple form:
\begin{equation}\label{eq:TwoModes}
{\mathcal H}= \frac{\hbar\Omega}{2}\left(\hat{a}^\dagger \hat{b}+\hat{b}^\dagger 
\hat{a}\right)
+\hbar\kappa \left(\hat{a}^{\dagger 2}\hat{a}^2+\hat{b}^{\dagger 2}\hat{b}^2\right)
\end{equation}
where $\hat{a},\hat{b}$ annihilate a particle in modes $u_a$ and $u_b$, $\kappa$ characterizes
the strength of the atomic interactions inside each condensate and
$\Omega$ is the Rabi coupling amplitude between the two condensates.
Here we have restricted for simplicity to the case where (i) the condensates 
have identical
interaction properties, (ii) the interactions between atoms
in different wells are negligible, and (iii) the Rabi coupling is resonant. 
The most general two-mode case could be treated along the same lines.

The direct numerical solution of the Schr\"odinger equation is performed in a basis
of Fock states $|n_a,n_b\rangle$ with $n_{a,b}$ particles in modes $u_a,u_b$.
The numerical integration is simplified by the fact that $n_a+n_b$ is a
quantity conserved by the Hamiltonian evolution.
We start with a state in which all atoms are in mode $u_b$, either in a Fock state
$|n_a=0,n_b=N\rangle$ (for the Fock state simulations)
or in a coherent state $\propto \exp(N^{1/2}\hat{b}^\dagger)
|0,0\rangle$ (for the coherent state simulations). We watch 
the time evolution of the mean fraction of particles in mode $u_a$, 
$ p_a\equiv\langle {\hat{a}^\dagger \hat{a}}\rangle /N$.

Mean-field theory (the Gross-Pitaevskii equation), valid in the limit $N\gg 1$
with a fixed $\kappa N/\Omega$ \cite{U1_est_content},
predicts periodic oscillations of $\langle {\hat{a}^\dagger \hat{a}}\rangle/N$; 
the peak-to-peak amplitude of the oscillations is equal to unity
if $\kappa N/\Omega < 1$, and is smaller than one otherwise \cite{TwoModes}.
In the exact solution the oscillations are no longer periodic due to
emergence of incommensurable frequencies in the spectrum of $H$.

\begin{figure}[p]
\epsfxsize=14cm \centerline{\epsfbox{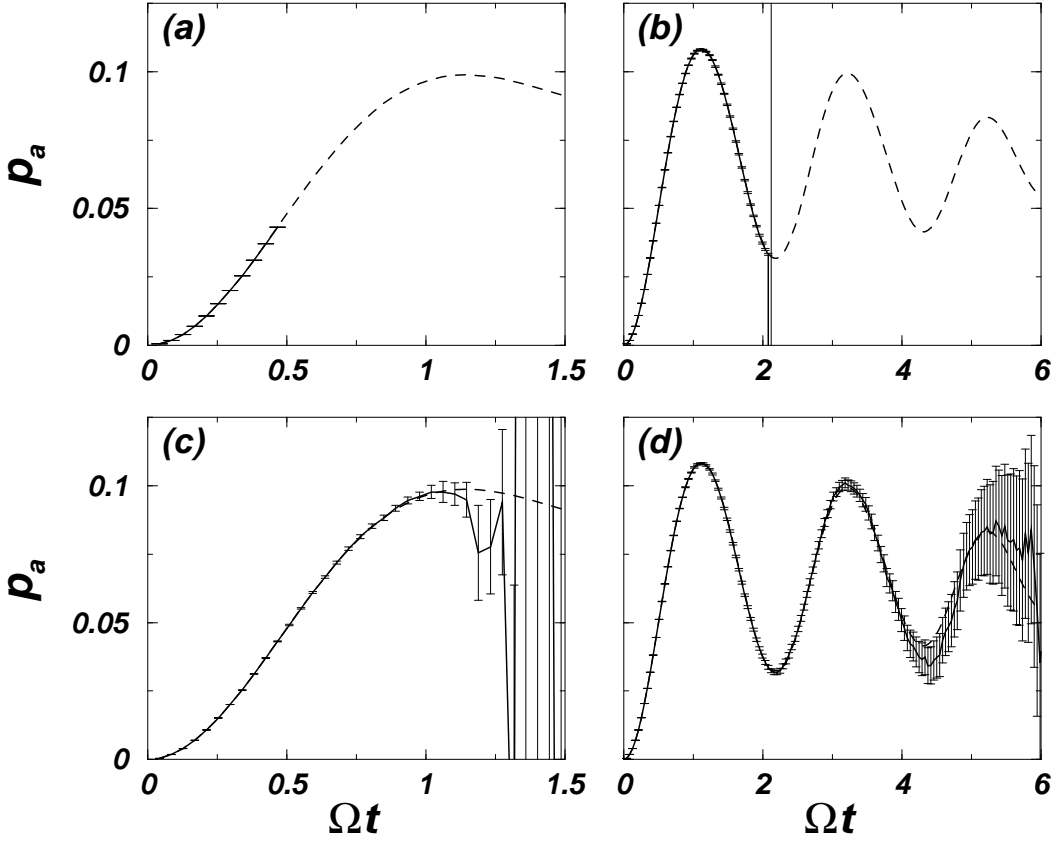}}
\caption{\small In the two-mode model mean fraction of atoms in the mode $u_a$ as function
of time, obtained with (a) the positive $P$-representation, (b) the Fock state simulation
with constant trace, (c) the simple simulation with coherent states,
and (d) the simple simulation with Fock states. The solid line represents
the average over ${\cal N}=2\times 10^5$ simulations, with corresponding error
bars. The dashed line is the direct numerical solution of the Schr\"odinger
equation. The number of atoms is $N=17$, initially all in mode $u_b$.
The interaction constant is $\kappa = 0.1\Omega$.
The time step used in the numerical stochastic
calculation is $\Omega dt=10^{-3}$.
The calculations in (a) and (b) have been stopped after the 
divergence of one realization. 
}\label{fig:TM}
\end{figure}

In the simulation method we evolve sets of two complex numbers
representing the amplitudes 
of the functions $\phi_1(x)$ and $\phi_2(x)$ on the modes $u_{a,b}(x)$
(plus the $\Pi$ coefficient in the coherent state case). 
The results are presented in figure \ref{fig:TM}
for $N=17$ particles and $\kappa/\Omega=0.1$, together
with the result of the direct integration of the Schr\"odinger equation.

The first row in the figure concerns the constant trace simulations.
Figure \ref{fig:TM}a shows results of the simulation based
on the positive $P$-representation, that
is the constant trace simulation with coherent states.
As well known \cite{Christ} this scheme leads to divergences of the norm 
$||\phi_1||\, ||\phi_2||$ for some realizations of the simulation. 
We have cut the plot in figure
\ref{fig:TM} at the first divergence.
The same type of divergences occurs in the constant trace simulation 
with Fock states (figure \ref{fig:TM}b). Note however that the characteristic time for the first
divergence to occur is somewhat longer. We have checked for these constant trace simulations
that the probability
distribution of $||\phi_1||\, ||\phi_2||$ broadens with time, eventually getting a power law tail.
The corresponding exponent $\alpha$  decreases in time below the critical 
value $\alpha_{\rm crit}=3$
for which the variance of $||\phi_1||\,||\phi_2||$ becomes infinite.
This scenario is identical to the one found with the positive $P$-representation \cite{Christ}.

The simple simulation schemes plotted on the second row
of figure \ref{fig:TM} provide results which are at all time in agreement
with the direct integration within the error bars. Contrarily to the constant
trace schemes we do not observe finite time divergences in the simple schemes.
For a given evolution time
we have checked that the error bars scale as $1/\sqrt{\cal N}$ where 
${\cal N}$ is the number of stochastic realizations. For a given ${\cal N}$
we found that the error bars increase quasi-exponentially with time.

\begin{figure}[p]
\epsfxsize=14cm \centerline{\epsfbox{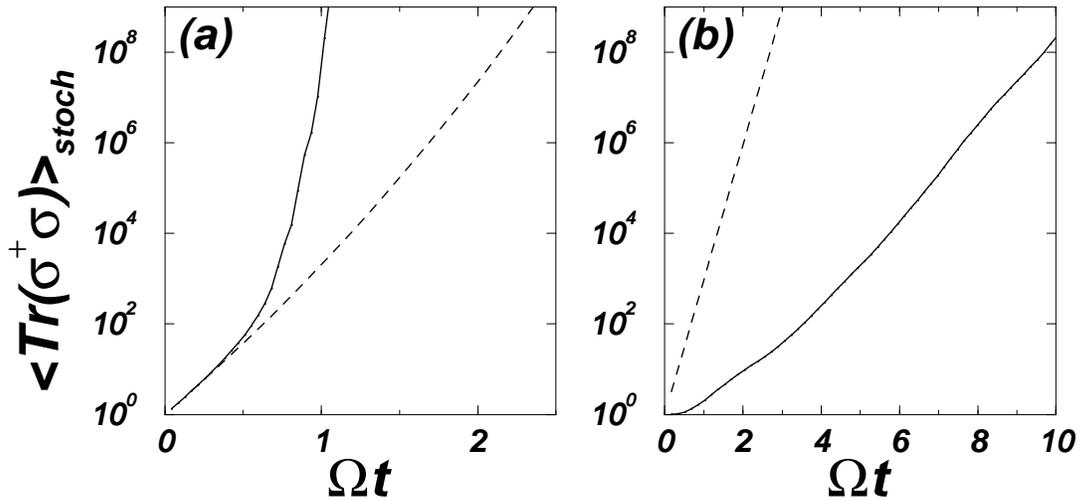}}
\caption{\small Statistical error on the $N-$body density matrix for the two-mode model:
(a) \lq simple' scheme with coherent states and (b)
\lq simple' scheme with Fock states.
The solid line is the numerical result of the simulations.
The dashed lines in (a) and (b) correspond respectively
to the lower and upper bounds Eq.(\ref{eq:min_Delta}) and Eq.(\ref{eq:maj_Delta}).
The parameters are the same as in figure \ref{fig:TM}.}\label{fig:trace}
\end{figure}

The noise in the simple simulation schemes is investigated in more details
in figure \ref{fig:trace} which shows the error estimator $
\Big\langle\mbox{Tr}[\sigma^\dagger\sigma]\Big\rangle_{\rm stoch}$ 
as function of time, for coherent states in figure \ref{fig:trace}a
and for Fock states in figure \ref{fig:trace}b.
The coherent state result confirms the prediction Eq.(\ref{eq:min_Delta}).
The Fock state result is found to be notably smaller than the upper bound
Eq.(\ref{eq:maj_Delta}). This is due to the fact that the terms proportional
to $\langle\phi_\alpha,\phi_\alpha|V|\phi_\alpha,\phi_\alpha\rangle$ in Eq.(\ref{eq:dtrSF})
are not negligible as compared to the term $V(0)$.
We have checked these conclusions for various values of $N$ and $\kappa/\Omega$.

\begin{figure}[htb]
\epsfxsize=14cm \centerline{\epsfbox{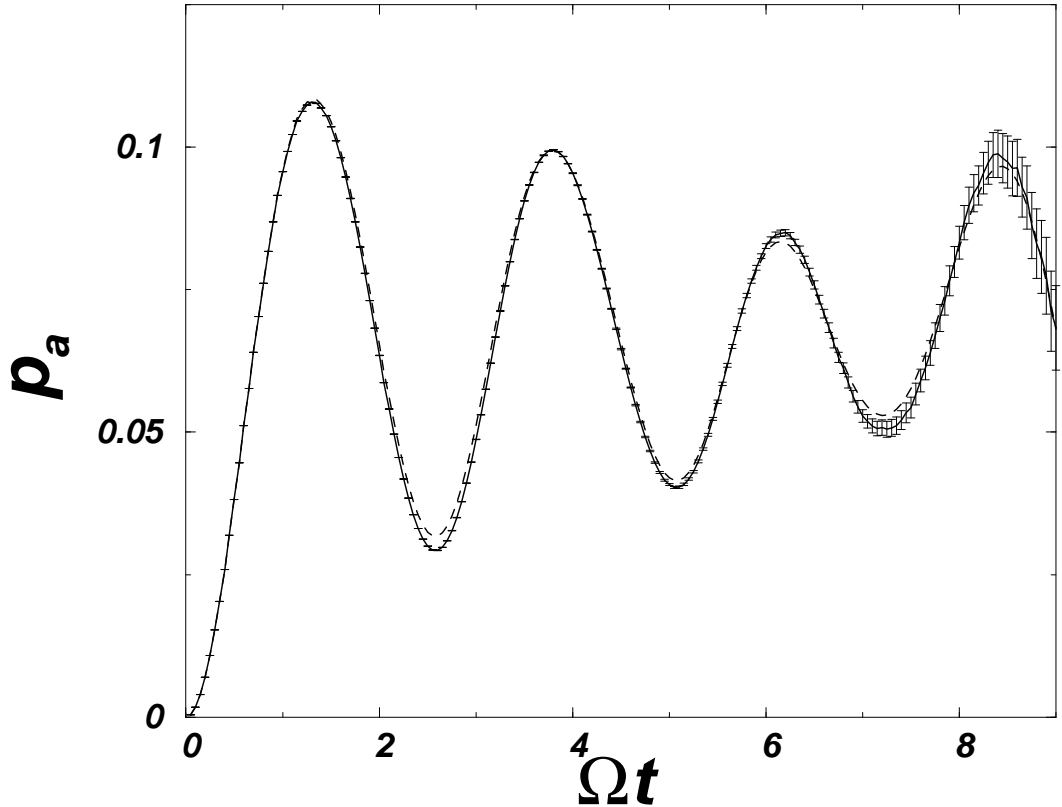}}
\caption{\small Fraction of atoms in mode $u_a$ in the two-mode model, for the parameters
of figure \ref{fig:TM}. 
The dashed line is the direct numerical solution of
the Schr\"odinger equation. 
The solid line with error bars is the result of
the \lq simple' scheme simulation with Fock states with ${\cal N}=10^8$
realizations. 
To keep a reasonable computation time with such a large 
value of ${\cal N}$ we have increased the time step in the numerical stochastic
integration
by a factor 25 with respect to figure \ref{fig:TM}. This explains the small systematic
deviation of the simulation result from the exact result visible for example
at time $\Omega t =2.6$.
The quantum phenomenon of collapse and revival
of the oscillation amplitude clearly apparent on the exact result is
well reproduced by the simulation.}\label{fig:revival}
\end{figure}

For a large number of particles
it is known \cite{revival} that the oscillations of $\langle \hat{a}^\dagger \hat{a}\rangle$
experience a collapse followed by revivals. These revivals are purely quantum
phenomena for the field dynamics and they cannot be obtained in classical field
approximation such as the Gross-Pitaevskii equation.
We expect to see a precursor of this phenomenon even for the small number of particles
$N=17$.
As the simple scheme simulation with Fock states is the most efficient of the four
schemes for the  investigation of the long time limit,
we have pushed it to the time at which a \lq\lq revival" can be seen,
as shown in figure \ref{fig:revival}. This figure is obtained with 
${\cal N} = 10^8$ simulations.

\section{Stochastic approach for a one-dimensional Bose gas}\label{sec:1D}
The interacting Bose gas is in general a multi-mode
problem, and the simulation schemes may have in this case a behavior  different 
from the one in a few-mode model such as in \S\ref{sec:MC}.
We have therefore investigated a model for a one-dimensional Bose gas.
The atoms are confined in a harmonic trap with an oscillation frequency
$\omega$. They experience binary interactions with a Gaussian interaction
potential of strength $g$ and range $b$:
\begin{equation}
V(x-y) = \frac{g}{(2\pi)^{1/2}b} \exp\left[-(x-y)^2/(2b^2)\right].
\end{equation}
At time $t=0$ all the atoms are in the same normalized 
state $\phi$ solution of the time independent Gross-Pitaevskii equation
\begin{equation}\label{eq:GPS}
\mu\phi(x)=-\frac{\hbar^2}{2m}\frac{d^2\phi}{dx^2}+
\frac{1}{2}m\omega^2 x^2\phi(x)+(N-1)\int\!dx'\;V(x-x')
|\phi(x')|^2\phi(x).
\end{equation}
At time $t=0^+$ the trap frequency is suddenly increased by a factor two, 
which induces a breathing of the cloud \cite{breathing1,breathing2,breathing3}.

\begin{figure}[p]
\epsfxsize=12cm \centerline{\epsfbox{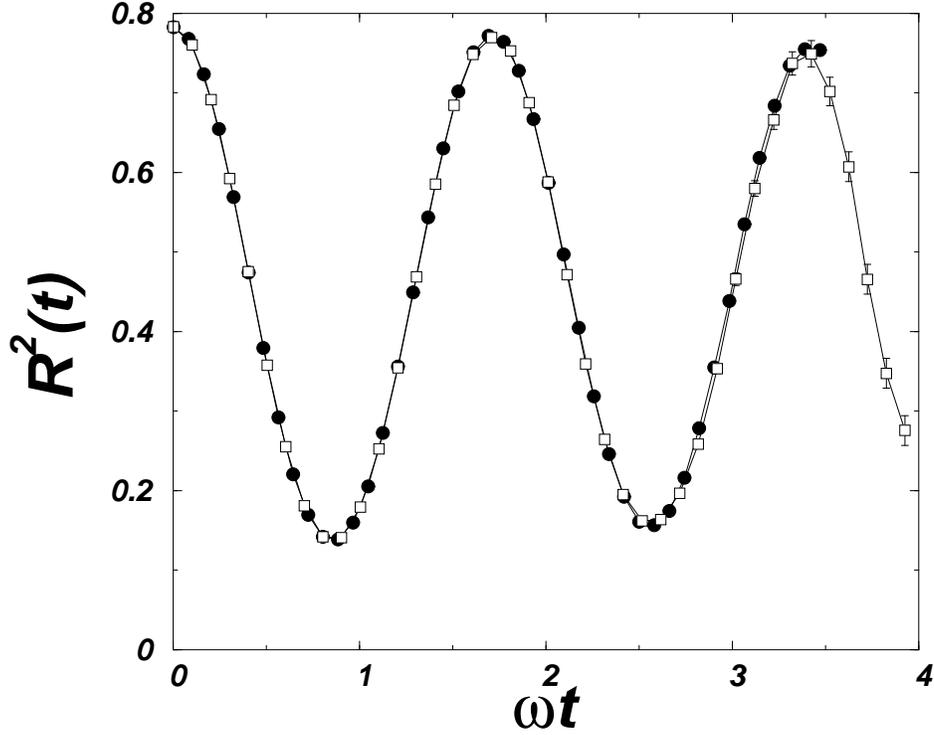}}
\caption{\small Mean squared spatial width $R^2$ of a harmonically confined
cloud of $N=10$ atoms 
as function of time. The breathing of the cloud
is induced by an abrupt change of the trap frequency from $\omega$ to $2\omega$.
The width $R$ is measured in units of the harmonic oscillator
length $a_{\rm ho}=(\hbar/(m\omega))^{1/2}$.
The interaction
potential is chosen such that $b=0.5 a_{\rm ho}$ and $g=0.4 \hbar\omega a_{\rm ho}$
leading to a chemical potential $\mu=1.7\hbar\omega$ in the Gross-Pitaevskii
equation Eq.(\ref{eq:GPS}). The calculation is performed on a spatial
grid with 32 points ranging from $- 6a_{\rm ho}$ to $+6a_{\rm ho}$
(with periodic boundary conditions). 
$\bullet$: constant trace simulation with ${\cal N} = 1000$
realizations. For $\omega t>3.5$ a divergence has occured for one of the realizations
and the calculation has been stopped.
$\Box$: simple scheme simulation with ${\cal N} = 40000$ realizations.
 }\label{fig:1D_largeur}
\end{figure}

\begin{figure}[p]
\epsfxsize=12cm \centerline{\epsfbox{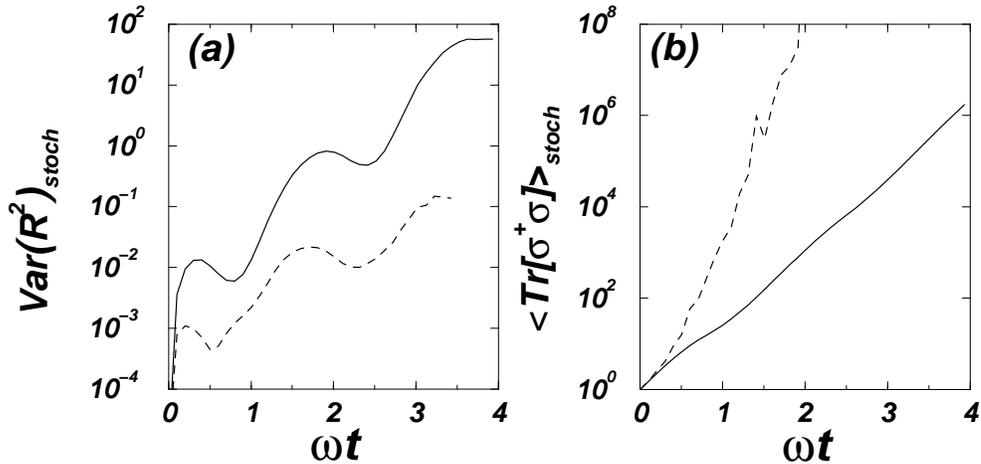}}
\caption{\small For the one-dimensional Bose gas in the conditions 
of figure \ref{fig:1D_largeur}, (a) stochastic variance of the size squared
of the cloud and (b) noise
on the $N-$body density matrix. Solid lines: simple scheme with Fock states.
Dashed lines: constant trace scheme with Fock states.}\label{fig:1D_bruit}
\end{figure}

This expected breathing is well reproduced by the numerical simulations.
The mean squared spatial width $R^2$ of the cloud as function of time is obtained
by taking $\hat{O}= \sum_{k=1}^N \hat{x}_k^2/N$ in Eq.(\ref{eq:StochMeanO}) where
$\hat{x}_k$ is the position operator of the $k-$th particle.
The quantity $R^2$ is shown in
figure \ref{fig:1D_largeur} for the simulation schemes with Fock states.
One recovers the key feature of the constant trace simulation,
that is a divergence of the norm $||\phi_1||\,||\phi_2||$ in finite time for
some realizations.
Before the occurrence of the first divergence the stochastic variance of the size squared of the
cloud, defined as
\begin{equation}
\mbox{Var}(R^2)_{\rm stoch} = \frac{1}{\cal N}\sum_{i=1}^{\cal N} \left[R_i^2(t)-R^2(t)\right]^2
\qquad \mbox{with} \qquad R_i^2(t) = {\cal R}e
\left[\langle N:\phi_2^{(i)}(t)|\hat{O}|N:\phi_1^{(i)}(t)\rangle\right]
\end{equation}
is notably smaller in the constant trace scheme than in the simple scheme, as shown in figure
\ref{fig:1D_bruit}a. This contrast between the two schemes
for the statistical error on one-body observables was absent in the two-mode
model of \S\ref{sec:MC}. 

We have also investigated the noise on the $N-$body density matrix
characterized by $\Big\langle\mbox{Tr}[\sigma^\dagger\sigma]\Big\rangle_{\rm stoch}$
(see figure \ref{fig:1D_bruit}b).
As expected this error indicator is smaller with the simple scheme. For this simple
scheme it varies quasi exponentially with time with an exponent $\gamma
\simeq 4 \omega$, which is smaller by
a factor roughly 2 than the one of the
upper bound Eq.(\ref{eq:maj_Delta}). This difference is due to the fact
that the range $b$ of the interaction potential is chosen here of the same
order as the size $R$ of the cloud so that the terms
$\langle \phi_{\alpha},\phi_{\alpha}|V|\phi_{\alpha},\phi_{\alpha}\rangle$ 
neglected in the derivation of 
the upper bound are actually significant. We have checked for various ranges $b$
much smaller than $R$ that $\gamma$ then approaches
the upper bound $2N V(0)/\hbar$.

\section{Conclusion and perspectives}
In this paper we have investigated a general method to solve exactly
the $N-$body problem in the bosonic case. The principle of the approach is to
add to the usual mean-field Gross-Pitaevskii equation a fluctuating
term. We have determined the general conditions ensuring that the average
over all possible realizations of this stochastic equation reproduces
the exact $N-$body Schr\"odinger equation.

This idea already received a particular implementation in quantum optics,
in the frame of the positive $P$-representation. We recover here the scheme
based on the positive $P$-representation
as a particular case of a simulation evolving coherent states of the
bosonic field with the constraint that the trace of the density operator 
should remain exactly equal to unity for each single realization.
This provides a simple derivation of the stochastic evolution within this representation 
alternative to the usual one \cite{PositiveP} based on analyticity properties.

Among the many possible implementations of the general stochastic approach
we have also investigated schemes evolving Fock states (that is number states) 
rather than coherent states. This is well suited to situations where the
total number of particles is conserved. In particular we have identified a scheme
preserving exactly the trace of the density operator which is for number
states the counterpart of the one based on the positive $P$-representation. 

Schemes with constant trace are subject to divergence of the norm
of some realizations in finite time. This effect, already known in the context
of the positive $P$-representation \cite{Christ}, makes these schemes difficult
to use.

In order to overcome this divergence problem we have investigated schemes
in which the condition on the trace is relaxed.  We have chosen instead
to minimize the statistical spread on the $N-$body density matrix, which gave
rise to the \lq simple' schemes, either with coherent states or Fock states.
In this case the $N-$body density operator is obtained as a stochastic
average of dyadics such as $|\mbox{coh}:\bar{N}^{1/2}\phi_1\rangle
\langle \mbox{coh}:\bar{N}^{1/2}\phi_2|$ or 
$|N:\phi_1\rangle\langle N:\phi_2|$, where the evolutions of $\phi_1$ and $\phi_2$
are fully decoupled. The deterministic parts are given by Gross-Pitaevskii equations,
which preserves the norm of $\phi_{1,2}$, contrarily to the case of constant trace schemes. 
The decoupling between the evolutions of $\phi_1$ and $\phi_2$ allows
a reinterpretation of our representation of the $N-$body density operator.
If the initial density operator is given by
$\rho(t=0)= |N:\phi_0\rangle\langle N:\phi_0|$ it will be given at time $t$
by
\begin{equation}
\rho(t) = |\Psi(t)\rangle\langle\Psi(t)|
\end{equation}
with the $N-$particle state vector
\begin{equation}
|\Psi(t)\rangle = \mbox{lim}_{{\cal N}\rightarrow\infty} 
\frac{1}{\cal N} \sum_{j=1}^{\cal N}
|N:\phi^{(j)}(t)\rangle.
\end{equation}
In this expression $\phi^{(j)}$ are stochastic realizations 
with the initial condition $\phi^{(j)}(t=0)=\phi_0$.

The \lq simple' schemes have much better stability properties than the
constant
trace schemes: differently from the case of constant trace schemes,
the deterministic evolution of the \lq simple' schemes has a
Gross-Pitaevski form and thus conserves the norms
$||\phi_{1,2}||$. This condition, together with the upper bound
\mbox{$\zeta_{1,2}\leq V(0)dt||\phi_{1,2}||^2/\hbar$}
on the eigenvalues $\zeta_{1,2}$ of the noise covariance operator 
$\overline{dB_\alpha(x)dB_\alpha^*(x')}$, can
be used to prove that the stochastic
equations possess a finite, {\it non-exploding} solution
valid for all times (see \cite{Gardiner} in page 94, \cite{McKean} in
\S 4.5).

We have numerically applied the simulation schemes to a two-mode model and to a one-dimensional
Bose gas. In both cases we found that the constant trace schemes lead to some diverging 
realizations, while the simple schemes lead to a statistical spread on the
$N-$body density operator increasing exponentially with time
with an exponent $\gamma\propto N V(0)$.
The simple schemes are therefore not well suited to determine small deviations
from the mean-field approximation in the large $N$ limit but can be more efficiently
applied to systems with a small number of particles, such as small atomic clouds
tightly trapped at the nodes or antinodes of an optical lattice.

In the one-dimensional numerical example of this paper we have
presented results for a simple one-body observable, the size of the atomic
cloud. We have actually
extended the calculations to more elaborate observables such as the first order and the second
order correlation functions of the field. We have not presented the results here as
the initial state of the gas was taken to be a (not very physical) Hartree-Fock state.
We are presently working on the possibility to generate a more realistic initial state
such as a thermal equilibrium for the gas, by extending our stochastic approach to
evolution in imaginary time.

This work has several possible perspectives of extension. One can first use
as building block a more sophisticated ansatz than the Hartree-Fock state 
$|N:\phi\rangle$,
such as Bogoliubov vacua (that is squeezed states of the atomic field)
or a multimode Hartree-Fock ansatz (that is an arbitrary coherent superposition
of number states in several adjustable modes of the field).
One can also look for {\sl approximate} rather than {\sl exact} 
stochastic solutions to the $N-$body
problem but that would be better than mean-field approaches in some given situations.
We hope to address some of these perspectives in the near future.

%\begin{references}


\begin{thebibliography}{9}
\bibitem{nonposit} One can show that the distribution ${\cal P}(\phi)$
evolves according to a functional Fokker-Planck equation with a non-positive
diffusion coefficient. This cannot be simulated by a Brownian motion
of $\phi$. In quantum optics this problem is known to
occur in the Glauber-Sudarshan $P$-representation of the density
matrix. See e.g. \cite{Gardiner}, chap. 10, \S 7.
\end{thebibliography}
\end{document}